# Carrier Localization in Pnictogen-Based Chalcohalides from Defect-Bound Hot Polarons

Xiaoyu Guo[1#], Junzhi Ye[1,2#*], Cibrán Lopez Alvarez[3,4], Maciej Oskar Liedke[5], Maik Butterling[6], Mutibah Alanazi[7], Yi-Teng Huang[1,8], Jiajie Wu[9], Zhilong Zhang[9], Lars Van Turnhout[10], Yorrick Boeije[10], Bofeng Xue[10], Qingyu Wang[7], Hugh Lohan[1], Seán R. Kavanagh[11], Andreas Wagner[5], Eric Hirschmann[5], Robert A. Taylor[7], Akshay Rao[10], Edgardo Saucedo[4,12], Claudio Cazorla[3,4,13], Robert L. Z. Hoye[1*]

1. Inorganic Chemistry Laboratory, University of Oxford, South Parks Road, Oxford OX1 3QR, United Kingdom
2. Institute of Polymer Optoelectronic Materials and Devices, Guangdong Basic Research Centre of Excellence for Energy & Information Polymer Materials, State Key Laboratory of Luminescent Materials and Devices, School of Materials Science and Engineering, South China University of Technology, Guangzhou 510640, China
3. Departament de Física, Universitat Politècnica de Catalunya, Campus Diagonal-Besòs, Av. Eduard Maristany 10–14, Barcelona 08019, Spain
4. Barcelona Research Center in Multiscale Science and Engineering, Universitat Politècnica de Catalunya, Institució Catalana de Recerca i Estudis Avançats (ICREA), Passeig Lluís Companys 23, 08010 Barcelona, Spain
5. Helmholtz-Zentrum Dresden – Rossendorf, Institute of Radiation Physics, Bautzner Landstr. 400, 01328 Dresden, Germany
6. Reactor Institute Delft, Department of Radiation Science and Technology, Faculty of Applied Sciences, Delft University of Technology, Mekelweg 15, NL-2629 JB Delft, The Netherlands
7. Clarendon Laboratory, Department of Physics University of Oxford, Oxford, OX1 3PU, United Kingdom
8. Graduate Institute of Photonics and Optoelectronics and Department of Electrical Engineering, National Taiwan University, Taipei 10617, Taiwan
9. South China Advanced Institute for Soft Matter Science and Technology, School of Emergent Soft Matter, South China University of Technology, Guangzhou 510640, China
10. Cavendish Laboratory, University of Cambridge, 11880, Cambridge CB3 0HE, United Kingdom
11. Harvard University Center for the Environment, Cambridge, Massachusetts, 02138, USA
12. Departament d'Enginyeria Electrònica, Universitat Politècnica de Catalunya, 08034 Barcelona, Spain
13. Institució Catalana de Recerca i Estudis Avançats (ICREA), Passeig Lluís Companys 23, 08010 Barcelona, Spain

# These authors contributed equally to this work.

* Email: junzhiye1994@scut.edu.cn (J.Y.), robert.hoye@chem.ox.ac.uk (R. L. Z. H.)

**Abstract**

Pnictogen-based solar absorbers have gained prominence as promising nontoxic and stable alternatives to lead-halide perovskites (LHPs), but are severely limited by carrier localization, preventing their performance from approaching those of LHPs. Recent efforts have uncovered routes to overcome carrier localization, but these early efforts only considered intrinsic factors. Herein, we push beyond these limited early efforts, examining the role of defects, not only on cold carriers but also hot carriers. Focusing on the structurally one-dimensional pnictogen chalcohalide BiSBr, we find that whilst this material intrinsically does not exhibit carrier localization, vacancies introduced during synthesis or post-treatment lead to pronounced extrinsic self-trapping via the formation of defect-bound hot polarons—excited charge-carriers strongly coupled to local defect-induced vibrational modes. These above-gap defect states divert hot carriers from cooling to the band edge, thus depleting the mobile carrier population. Our findings establish the key role of defect-bound hot polarons in mediating extrinsic localization and offer new mechanistic insights into the interplay between defects, lattice coupling, and excited-state charge-carrier transport, which are critical to designing efficient perovskite-inspired solar absorbers.

## Main

The rapid rise of lead-halide perovskites (LHPs) in photovoltaics has prompted a substantial effort to develop solar absorbers capable of replicating the performance of these materials, whilst overcoming their toxicity and stability limitations[1,2]. These perovskite-inspired materials (PIMs) have heavily focussed on compounds based on heavy pnictogen cations ($Bi^{3+}$, $Sb^{3+}$), which share key features of the electronic structure of LHPs at band extrema that are conducive towards defect tolerance[2,3]. Although many pnictogen-based PIMs exhibit greater environmental and thermal stability than LHPs, and are fully free from toxic, regulated elements, their performance in photovoltaics has fallen well short of LHP solar cells[2]. A critical limiting factor is the low charge-carrier mobilities of these PIMs, which restricts the diffusion lengths and charge-collection efficiency of these materials in devices[4-7]. This is caused by carrier localization, which has been so widely found in heavy pnictogen-based PIMs, that it is being labelled a hallmark of these materials[8-13].

Carrier localization occurs when the wavefunction of an electron, hole or exciton is constrained to within a unit cell, forming a small polaron or self-trapped exciton with severely reduced charge-carrier mobility. This arises from strong coupling between charge-carriers and longitudinal optical (LO) phonons (Fröhlich coupling) or acoustic phonons, and is characterized by a rapid decrease in charge-carrier mobility on the order of a picosecond after photoexcitation[14,15]. Some of the underlying causes behind the wide occurrence of carrier localization among heavy pnictogen-based PIMs are their low electronic dimensionality (*e.g.* $NaBiS_2$[16,17]), as well as high acoustic deformation potential (*e.g.*, for $Cs_2AgSbBr_6$ and $Cs_2AgBiBr_6$[7,9]). Recently, the field started to find exceptions to this prevalence of carrier localization among pnictogen-based PIMs. An early example was BiOI, where we found delocalized charge-carriers through both computations and experiment[18,19], followed by $CuSbSe_2$, where we also found polarons to be large[20]. In the latter case, we rationalized that this deviation from the norm for PIMs arises due to 1) regular free volume in the structure (i.e., gaps between slabs of material in this layered compound) relaxing distortions from the propagation of acoustic waves, thus lowering deformation potentials, 2) quasi-bonding between layers leading to increased electronic dimensionality, and 3) low ionic contributions to the dielectric constants due to the low bandgaps and low Born effective charges, which then lower the Fröhlich coupling constant. However, these are all intrinsic factors, whereas point defects (e.g., vacancies, anti-sites and interstitials) inevitably form in materials at room temperature. The role of these defects on carrier localization in PIMs is not well understood.

Here, we refer to extrinsic carrier localization as the restriction of the charge-carrier wavefunction to within a unit cell as a result of defect-bound polarons, resulting from strong carrier coupling to defect-induced local vibrational modes. This contrasts to intrinsic carrier localization driven solely by coupling to vibrational modes in the host structure[21,22]. Furthermore, the discussion on carrier localization in PIMs thus far has only focussed on cold carriers, whereas defects and polarons also influence hot carrier cooling, with significant consequences on the efficiency limit of solar absorbers[23].

To resolve these pressing questions, we focus on BiSBr as an exemplar pnictogen-based PIM. This is a highly novel solar absorber, having only recently been demonstrated for photovoltaics[4,24], with little understood about its photophysics. This material is structurally one-dimensional, with Bi $6s^2$ lone pairs directionally oriented towards Br from neighbouring ribbons. Thus, from what we recently established by investigating $CuSbSe_2$, we would expect BiSBr to also intrinsically exhibit band-like transport due to its regular free volume, along with quasi-bonding between ribbons. At the same time, BiSBr has complex defect chemistry, with transition levels forming both within the bandgap, as well as within conduction band[25]. BiSBr therefore presents an ideal platform to understand the role of defects on the localization of hot carriers, and the consequences on cold carrier kinetics. With this aim, we examined this material through a combination of simulations of defects and polarons, along with advanced spectroscopic measurements, including positron annihilation spectroscopy and pump-push-probe spectroscopy. We establish a direct link between vacancy clusters and defect-bound hot polaron states—excited carriers strongly coupled to local lattice vibrations. This work not only identifies a previously overlooked localization mechanism, but also provides a framework for explaining how defects limit mobility and charge-carrier extraction efficiency. The findings highlight the key role of defect states in governing device performance through hot carriers, and contributes a new set of design principles for developing efficient solar absorbers.

**Results**

***BiSBr intrinsically exhibits band-like transport***

To examine the effects of defects on carrier localization in BiSBr, we prepared samples using two different synthesis methods (bulk thin films and nanocrystals, NCs) with the intention of tuning the defect density. Bulk thin films were prepared by solution processing in a nitrogen-filled glovebox, while NCs were prepared by hot injection in a Schlenk line, and purified NCs were drop cast to prepare films (detailed in Methods). As shown in **Supplementary Fig. 1**, these two methods yielded crystals with markedly different dimensions: the NCs (average length 83±8 nm) are substantially smaller than the bulk crystals (5.7±0.4 μm), resulting in a much larger surface-to-volume ratio, which could lead to a higher density of surface defects from unsaturated surface bonds and structural relaxation at the crystal boundaries (discussed quantitatively below). **Fig. 1a** shows the optical absorption spectra of both films, which exhibit similar bandgaps. We also confirm the samples to be phase-pure, with no phase impurities or impurity elements for either sample (**Fig. 1b** and **Supplementary Fig. 2-3**).

Despite the samples having the same structure, we found a marked difference in carrier localization. Optical pump terahertz probe (OPTP) spectroscopy showed that the NC sample exhibits a rapid decrease in photoconductivity within the first ps after photoexcitation, whereas the bulk film decayed in photoconductivity much more slowly (**Fig. 1c**; details in **Supplementary Note 1** and **Supplementary Fig. 4)**. The rapid ps decay in photoconductivity is too fast to realistically arise from defect trapping, and is mostly due to a reduction in mobility due to carrier localization[14]. Comparing the times taken for the OPTP signal to decay to half of the original peak value ($t_{1/2}$), the NCs have a $t_{1/2}$ <0.35 ps, which is comparable to PIMs that are known to exhibit carrier localization ($NaBiS_2$ with $t_{1/2}$ ~ 0.5 ps[26], $Cs_2AgBiBr_6$ with $t_{1/2}$ 1–2 ps[27,28]), as detailed in **Supplementary Table 1**. On the other hand, bulk thin films of BiSBr have $t_{1/2}$ > 8.5 ps, longer even than $CuSbSe_2$ ($t_{1/2}$ 6.7 ps), which we previously established to have band-like transport[18]. Additionally, since the NC sample does not show strong excitonic behaviour (exciton binding energy in the range of 6–29 meV from Elliott model fitting and first-principles calculations, **Supplementary Fig. 5**), the rapid drop of photoconductivity in the NC samples is unlikely to be due to exciton formation. This strongly hints at the determining role of defects on carrier localization.

To understand carrier localization in BiSBr in more detail, we first examined whether this process would intrinsically occur when no defects are present. To do this, we analysed the Fermi isosurfaces and acoustic deformation potentials to evaluate the extent of electronic anisotropy and electron-phonon coupling, complemented with polaron ionization energy first-principles calculations. Our results are consistent with expectations based on our prior work with $CuSbSe_2$[20]. The acoustic deformation potentials of BiSBr are low (1.67 eV for the VBM and 1.54 eV for the CBM; **Supplementary Table 2**), consistent with the presence of regular free volume in this structurally-1D material. Furthermore, our first-principles calculations of the Fermi isosurfaces 0.1 eV from the band extrema revealed the electronic dimensionality to be 2D in the lower conduction band, and 3D in the upper valence band (**Fig. 1d**). This is consistent with expectations based on the quasi-bonding that would occur between ribbons in BiSBr. We would therefore expect band-like transport to occur, and this is indeed what we found experimentally from BiSBr thin films.

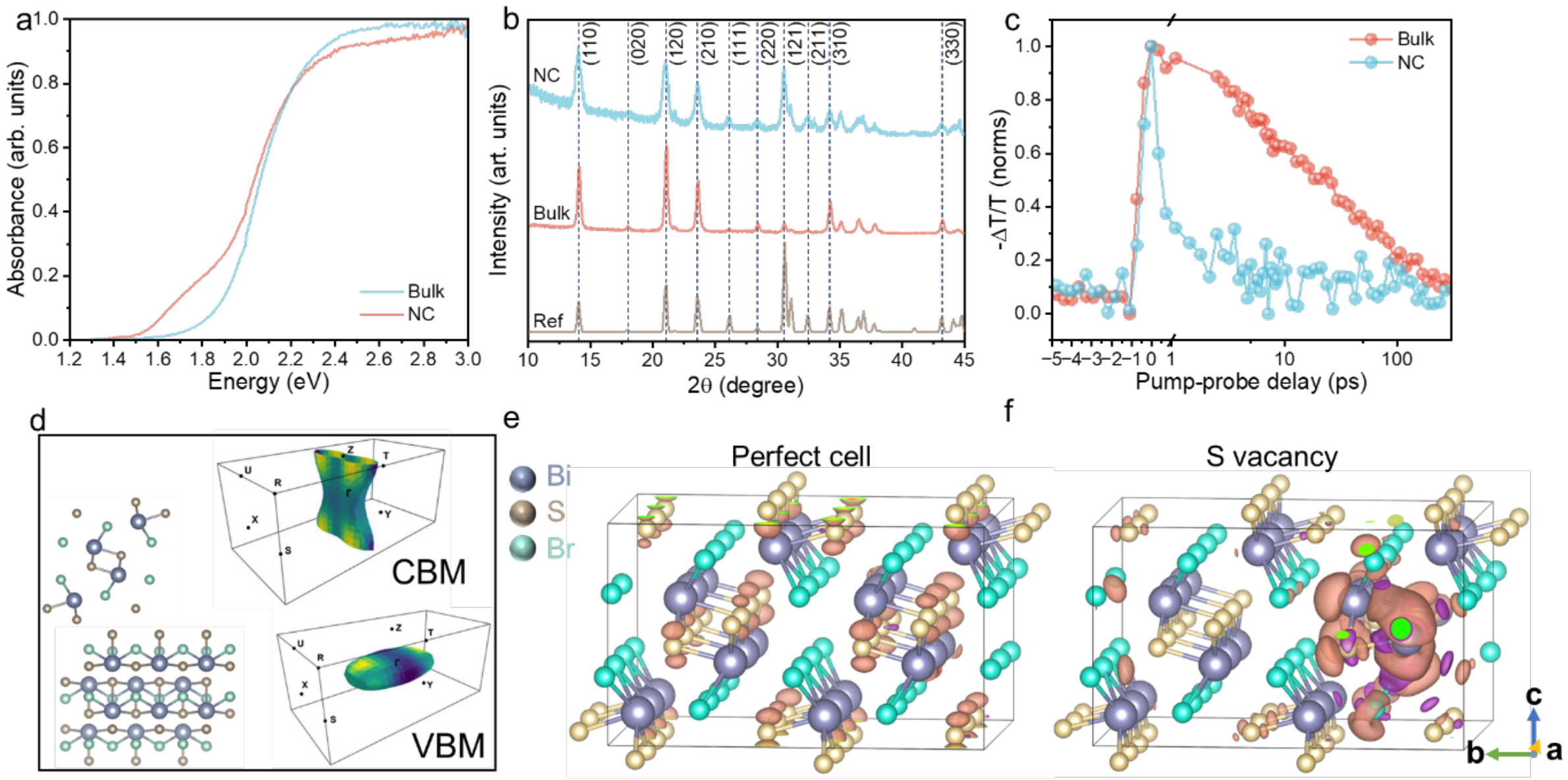

**Fig. 1 | Carrier localization takes places in BiSBr only with the introduction of defects. a,** Normalized absorption spectra of bulk and NC thin films of BiSBr, measured by photothermal deflection spectroscopy. **b,** X-ray diffraction (XRD) patterns of bulk and NC BiSBr thin films compared to the reference pattern (space group *Pnma*, ICSD database, Coll. Code: 31389). **c,** Optical pump terahertz probe (OPTP) spectra of bulk and NC BiSBr. **d,** Fermi iso-surfaces 0.1 eV below the valence band maximum (VBM; bottom) and above the conduction band minimum (CBM; top). Electron (brown regions) and hole (purple regions) charge density distributions in a **e,** pristine BiSBr supercell and **f,** defective BiSBr supercell containing a sulfur vacancy. In part **e**, the hole wavefunctions are not visible from this perspective, but are delocalized across the entire supercell, as are the electron wavefunctions. By contrast, in the defective supercell, the electron and hole wavefunctions become localized to the point defect.

***Carrier localization occurs in BiSBr when defects are present***

To understand how defects influence the nature of charge-carriers in BiSBr, we first modified our polaron first-principles calculations by introducing a sulfur vacancy ($V_S$) into the supercell (**Fig. 1e**), which was distorted slightly in order to search for its most stable configuration at each charge state. As shown in **Fig. 1f**, the presence of $V_S$ led to strong lattice distortion and electron localisation. In the –2 charge state ($V_S^{-2}$), one neighbouring bismuth atom next to the vacancy underwent a pronounced displacement, breaking bonds with surrounding bromine atoms and one of the two sulfur atoms. Bond breaking can be seen from the Bi–S bonds ($Bi_4$– $S_1$ in **Supplementary Fig. 6**) being elongated from 2.58 to 2.63 Å. This displacement drives the bismuth atom towards the vacancy, partially filling the empty site and trapping the excess electron as a small polaron (**Supplementary Fig. 7**). The change in bond lengths between different atoms in BiSBr before and after introducing a $V_S^{-2}$ defect is summarized in **Supplementary Table 3**. As a result, we can see that the electron and holes wavefunction become localized around the S vacancy (**Fig. 1f**). The defect polaron binding energy ranges from 0.26 eV for the negatively-charged $V_S$ defect, to 1.19 eV for the neutral $V_S$ defect (**Supplementary Table 4**). These values are similar to the reported polaron binding energies of $CsPbBr_3$ perovskite (0.45 eV)[29], and some metal oxides (1.0-2.6 eV)[30], both of which are considered to have small polarons. These first-principles calculations therefore demonstrate the feasibility of small polaron formation in BiSBr as a result of sulfur vacancies being present.

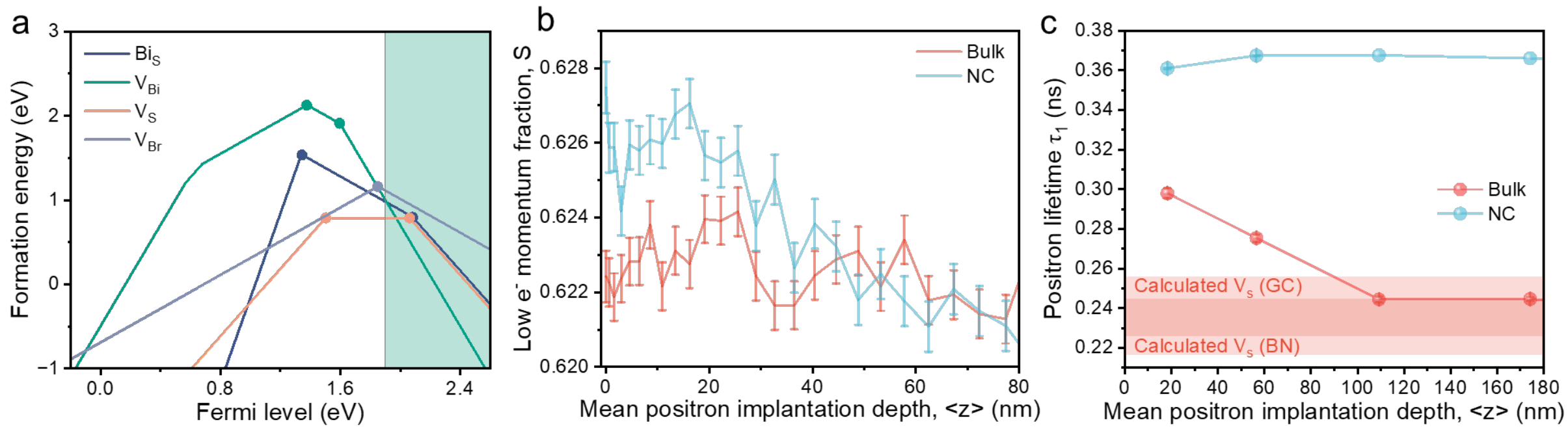

**Fig. 2 | Defect chemistry in bulk thin film and NC BiSBr. a,** Defect formation energies of BiSBr as a function of the Fermi level for four representative point defects in different charge states: $Bi_S$ (bismuth-on-sulfur antisite), $V_{Bi}$ (bismuth vacancy), $V_S$ (sulfur vacancy), and $V_{Br}$ (bromine vacancy). The shaded area represents area above conduction band edge. **b,** Low $e^-$ momentum fraction (*S*) and, **c,** First positron lifetime as a function of mean positron implantation depth for bulk and NC BiSBr extracted from PAS measurements.

In order to quantitatively compare the density of defects in the bulk thin films and NCs, we used positron annihilation spectroscopy (PAS), which measures electron momentum distribution at annihilation sites and the lifetime of positrons before annihilation with electrons, revealing information about open-volume defects, such as vacancies, as well as the local electron density[31,32]. The *S* parameter, acquired from Doppler broadening variable energy (DB-VE) PAS (see Supplementary Methods), quantifies the fraction of annihilations with low-momentum valence electrons, helping to distinguish between the environments of the bulk lattice and open-volume defects. **Fig. 2b** and **Supplementary Fig. 8** demonstrated that the *S* fraction, which scales with open volumes in materials, and the average positron annihilation lifetime (effective defect size) of the NC film is higher than the bulk film, indicating that there are higher concentrations of defects and/or larger sizes of defect clusters in the NCs. The longer positron lifetimes in the NC films indicate trapping at vacancy-type defects with reduced electron density, leading to a lower probability for annihilation. The larger *S* value of the NC thin films suggests increased annihilation with low-momentum electrons, which can occur when positrons are trapped in open-volume defects (*e.g.*, vacancies and voids), where the positron wavefunction overlaps less with high-momentum core electrons.

The comparison of the defect density in the bulk and NC samples can also be obtained from photoluminescence (PL) measurements. The effect of mid-bandgap defects on carrier recombination is discussed in **Supplementary Note 4**. PL decay measurements under varying excitation fluences (**Supplementary Fig. 12a**) reveal that bulk BiSBr exhibits shorter lifetimes at higher fluences, whereas NC BiSBr shows the opposite trend (detailed in **Supplementary Fig. 12b**). These opposing trends in PL lifetime with fluence are consistent with a higher density of mid-gap defects in the NC films compared to bulk thin films, as a greater fraction of charge-carriers are captured by defect states under high-fluence excitation in the NCs. To further probe the role of mid-bandgap defects, we also investigated the PL intensity with varying pump laser excitation energy, ranging from 460 nm to 590 nm wavelength (**Supplementary Note 5** and **Supplementary Fig. 13a&14**). The corresponding variation in sub-bandgap PL intensity (integrated from 660 to 700 nm wavelength) as a function of excess photon energy relative to the bandgap energy (1.9 eV, or approximately 650 nm wavelength) is presented in **Supplementary Figure 13b**. In the less-defective bulk films, the PL intensity remains relatively stable across different excitation energies, as more photoexcited hot carriers relax to the band edge before recombining. In contrast, NC films exhibit a pronounced reduction in PL intensity with increasing excitation energy, indicating that hot carriers are increasingly trapped by mid-bandgap defects and undergo additional non-radiative recombination[23,33].

Having quantitatively confirmed the higher defect density in the NCs, we next needed to understand what the likely dominant defects are. This can be established from the positron lifetime, since different types of defects offer different open volumes and electron densities for positron trapping, which changes the overlap between the positron and electron wavefunctions, and hence the annihilation rate. In order to determine the defects most likely to form in BiSBr, we calculated the defect diagram for this material from first-principles (**Fig. 2a**; details of calculations in Methods). The phase-stable region in chemical potential space for BiSBr is delimited by S-poor ($\mu_{Bi}$, $\mu_{S}$, $\mu_{Br}$) = (0, -0.60, -0.97) eV and Bi-poor conditions ($\mu_{Bi}$, $\mu_{S}$, $\mu_{Br}$) = (-0.90, 0, -0.67) eV. From this, the lowest formation energy point defects are S vacancies, Bi on S antisites ($Bi_S$) and Br vacancies ($V_{Br}$). Bi vacancies ($V_{Bi}$) are less likely to form, but exhibit a wide range of possible charge states. From XPS measurements, we confirm that both bulk and NC thin films of BiSBr are S-deficient (**Supplementary Table 6**), suggesting that $V_S$ are likely the most common point defect. Moreover, the bulk film exhibits a lower first positron lifetime ($\tau_1$) in **Fig. 2c**, which typically corresponds to positron annihilation at small point defects. The measured value falls between predictions made by the Boronski–Nieminen (BN) and Gradient Correction (GC) theoretical methods for sulfur monovacancies ($V_S$**,** **Supplementary Table 5** and **Supplementary Fig. 10-11**), based on the defective BiSBr crystal structure demonstrated in **Supplementary Fig. 9**. BN and GC denote the local-density and gradient-corrected electron-positron correlation schemes used in positron-lifetime DFT calculations, respectively. We note however, that the binding energy of positrons to neutral sulfur monovacancies is close to zero or negative. Hence, only weak positron trapping is possible, which would increase for negatively charged configurations. Therefore, bulk films may exhibit a combination of positron annihilation in the bulk and at $V_S$. In contrast, the NC film displays a larger $\tau_1$, which is consistent with the presence of larger defect clusters rather than isolated vacancies, and matches well with the calculated lifetimes for various defect cluster configurations (details in **Supplementary Notes 3** and **Supplementary Fig. 10-11**). Whilst we cannot definitively distinguish between the four different possible point defects based on the positron lifetime measurements, the S deficiency found from XPS suggests the prevalence of $V_S$ within the materials. Additionally, greater S deficiency in the NCs compared to the bulk thin films from XPS also suggests a higher defect concentration in the NC samples, consistent with larger defect clusters.

To provide further experimental evidence that defects lead to a change in carrier localization behaviour in BiSBr, we intentionally introduced defects into bulk thin films. This was achieved by heat treating the films for longer (from 10 min to 30 min), or at higher temperatures (from 250 °C to 270 °C) in a $N_2$-filled glovebox. From XPS measurements, we found that heat treatment led to a further reduction in S content, as well as a reduction in Br content, indicating an increase in defect clusters based especially on $V_S$ and $V_{Br}$. We confirmed that no changes in phase occurred in the bulk after heat treatment (**Supplementary Fig. 15a**). The impact of these additional defects on charge-carrier behaviour was studied using OPTP spectroscopy, as shown in **Supplementary Fig. 15b**. Both heat-treated films with reduced sulfur content exhibited a pronounced change in OPTP kinetics from the slow decay of the pristine thin film to significantly faster decay with intentional defect introduction, with kinetics now similar to that of the NC films. This is in strong agreement with carrier localization in BiSBr films being linked to the concentration of defects.

***Hot polaronic states***

We surprisingly found pronounced above-gap PL emission from both NC and bulk thin films. The emission spectra were broad and could be deconvoluted into several components (**Fig. 3a-b**), including not only at and below the bandgap of BiSBr (1.91 eV), but also above-gap luminescence centred at 2.22 eV and 2.45 eV. These above-gap peaks were especially pronounced in the NC films, pointing to hot carrier emission processes that persist beyond initial carrier thermalization. These observations raised a critical question of how carrier localization occurs through defect introduction, and what the mechanisms involved are. Addressing this is not straightforward, given that this BiSBr system has defect clusters rather than simple monovacancy defects. Nevertheless, we can probe the overall mechanisms involved through spectrally- and time-resolved PL and transient absorption spectroscopy (TAS) measurements.

To understand the origin of the above-gap states, we note that $V_S$ and $Bi_S$ both have transition levels forming within the lower conduction band of BiSBr (**Fig. 2a**). We verified through first-principles calculations of the projected density of states (PDOS) of BiSBr that $V_S$ indeed introduces states within the conduction band (**Supplementary Fig. 16**). Typically, hot electrons that relax into these states would then relax down to the CBM to form cold carriers, and we would then only have PL at or below the bandgap. That we have above-bandgap PL suggests that the relaxation pathways for these hot carriers are blocked, and carriers directly emit from relaxation from these hot states to the ground state.

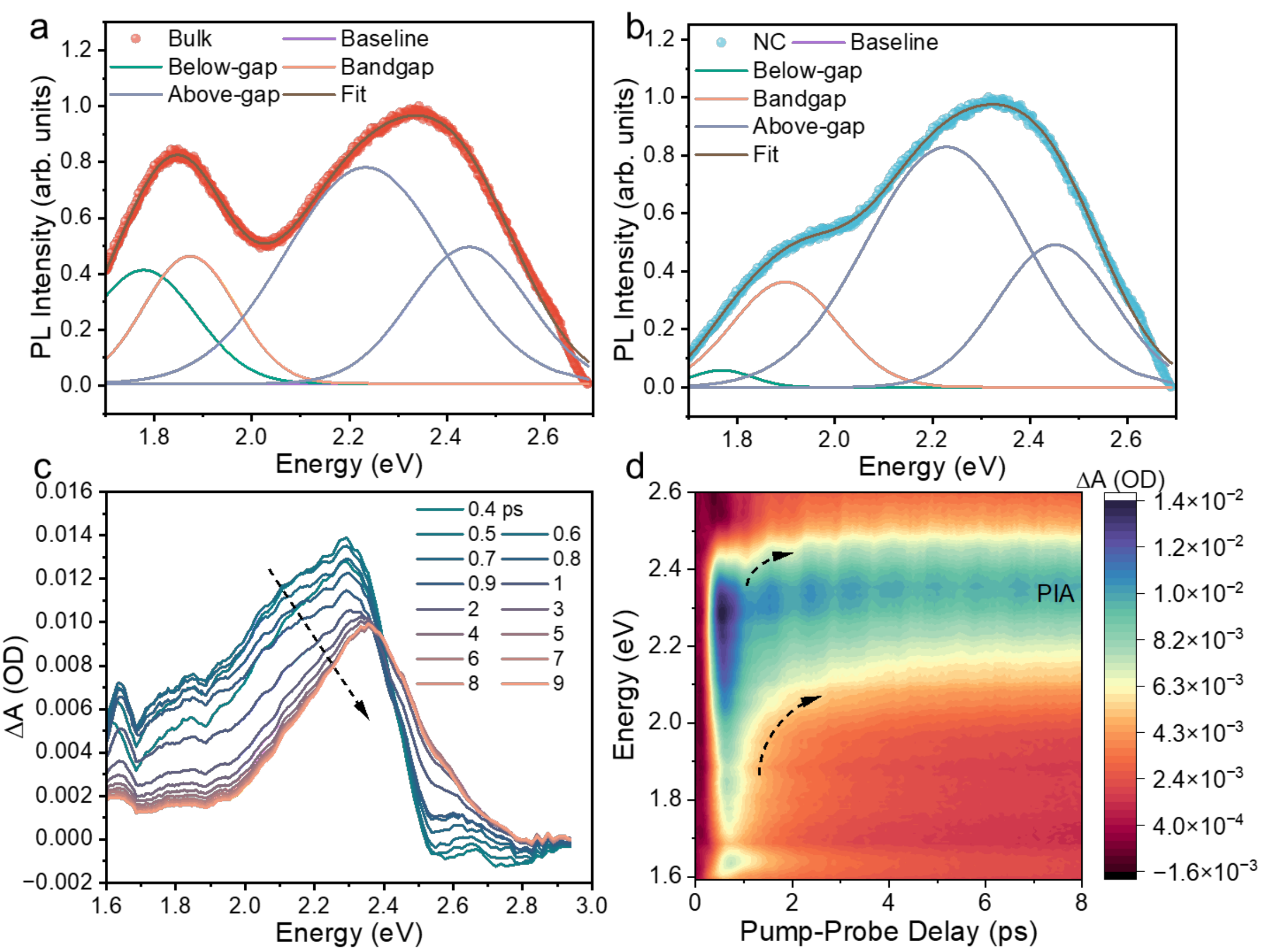

**Fig. 3 | Photoluminescence (PL) and transient absorption (TA) measurements of bulk and NC BiSBr thin films.** PL spectrum of **a,** bulk and **b,** NC films, excited with a 405 nm wavelength CW laser. Peak fitting shown. **c,** TA spectra of BiSBr colloidal NC solution collected within 9 ps after photoexcitation (excitation wavelength: 400 nm, repetition rate: 1 kHz, fluence: 40.66 μJ cm$^{-2}$). Positive differential absorption ($\Delta A$) corresponds to a photoinduced absorption (PIA). **d,** TA map of BiSBr NC solution showing the temporal evolution of the PIA. Refer to Supplementary Note 7 for our discussion of the origin of the PIA-dominated differential absorption.

To more directly probe the kinetics of hot carriers, we used TAS measurements (**Fig. 3c-d**). The TA spectra of the colloidal NC solution reveal a distinct blueshift in the photoinduced absorption (PIA; positive $\Delta A$) within the first 9 ps after excitation (**Fig. 3d**). This early-time spectral shift is a strong indication of hot small polaron formation[34]. We propose that following photoexcitation, electrons are initially injected into high-energy delocalized states. Within a few picoseconds, they undergo self-trapping due to strong electron–phonon coupling, forming small polarons that still retain excess energy above the band-edge. These hot small polarons occupy localized states slightly above the CBM, increasing the transition energy required for further excitation and leading to a blueshift in the TA spectrum. This trend is further visualized in the TA map in **Fig. 3d**, where the energy of the PIA peak increases during the initial few picoseconds following excitation. A comparison between BiSBr NCs and bulk BiSBr thin film is provided in **Supplementary Fig. 17**, showing a clear blueshift in the NC thin film, while the absorption energy remains unchanged in the bulk. These observations indicate that hot small polarons form in more defective BiSBr NCs, whereas such self-trapping is not evident in bulk BiSBr, possibly due to differences in defect density. Global fitting of the TAS measurements (**Supplementary Fig. 18**) involves constructing a kinetic model that represents possible carrier relaxation or decay pathways, and fitting this model to the entire time-wavelength dataset simultaneously. The proposed model that can achieve good global fitting of the two-pulse TA data is illustrated in **Supplementary Fig. 19**, where three relaxation steps are present at different energy levels, demonstrating that defects introduce intermediate states that intercept hot carriers before cooling to the band-edge.

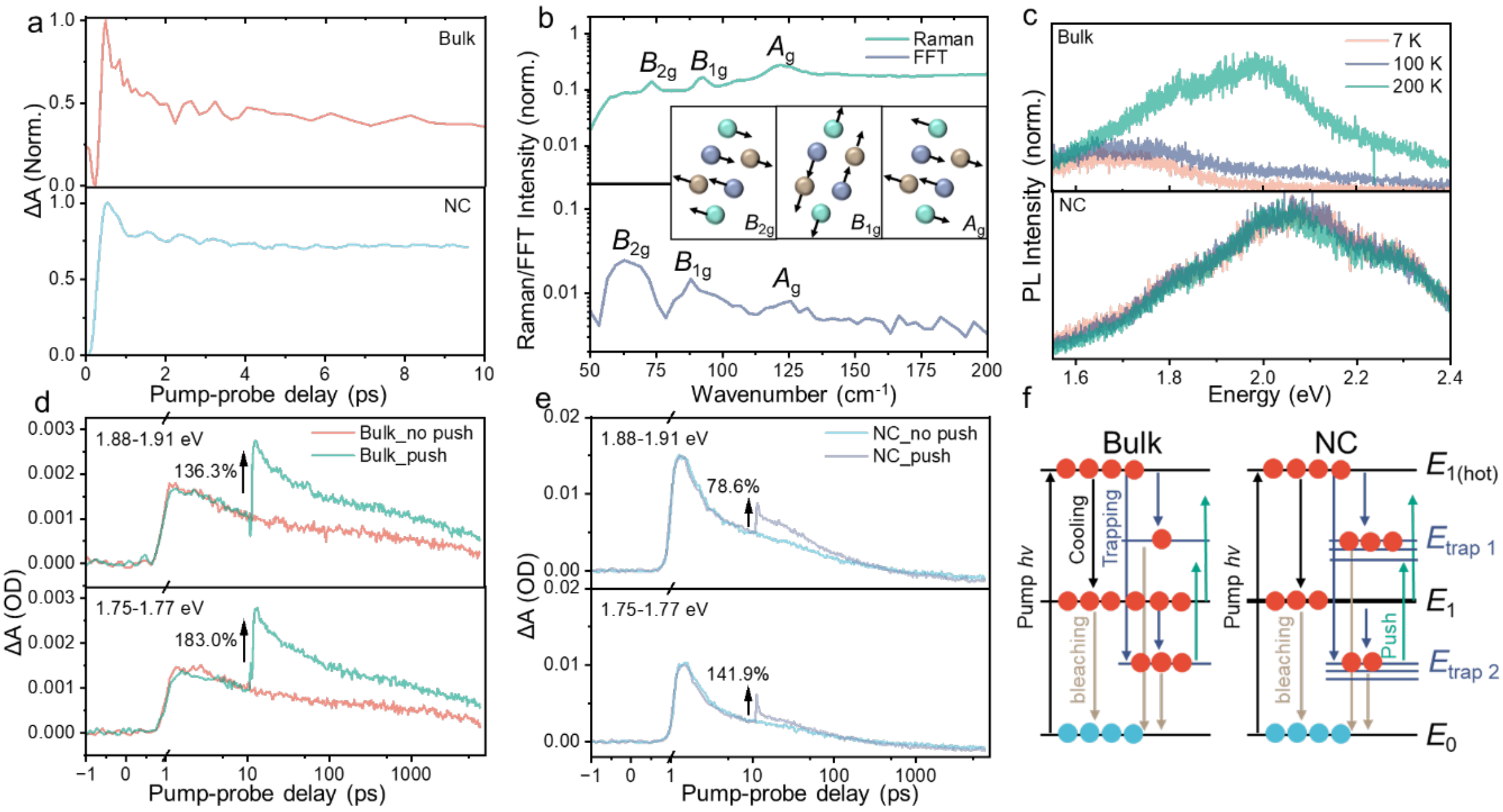

**Fig. 4 | Effect of above-bandgap defect transition levels on hot carriers. a,** Early time decay from two-pulse TA measurements probed at 2.38-2.25 eV for bulk and NC thin films with a 400 nm wavelength pump beam. Positive normalized $\Delta A$ corresponds to a photo-induced absorption (PIA). **b,** Comparison between the Raman spectrum of the NC film and its Fourier Transformed TA PIA. These vibrational modes have $B_{2g}$, $B_{1g}$ and $A_g$ symmetry for the $D_{2h}$[16] point group of BiSBr. **c,** PL spectra of bulk and NC films measured at 7 K, 100 K and 200 K. Kinetics of **d,** bulk and **e,** NC BiSBr from pump-push-probe TAS measurements probed at the band-edge (1.88-1.91 eV) and below the bandgap (1.75-1.77 eV) with and without the 1300 nm wavelength push beam (600 μW). The 400 nm wavelength pump has a fluence of 636.6 μJ cm$^{-2}$. The percentage changes in $\Delta A$ (OD) before and after the push beam are labelled in panel d and e. Details of the three-pulse TA is in **Supplementary Note 9**. **f,** Schematic of the transitions and relaxation pathways in the bulk and NC thin films. $E_{trap\ 1}$ corresponds to the

transition level for point defects within the conduction band, $E_1$ is the conduction band minimum, while $E_{\text{trap 2}}$ is the sub-gap trap state. $E_{1(\text{hot})}$ is the hot state charge-carriers are photo-excited into with the 400 nm wavelength pump beam, while $E_0$ is the ground state for the electron (valence band maximum).

To further understand the origin of small polaron formation, we analysed the early decay (0–10 ps) of the TA signal for bulk and NC film samples probed above the bandgap (2.25-2.38eV) (**Fig. 4a**). In the NC sample, a clear oscillation appears in the early decay signal. These oscillations are only present when BiSBr is photoexcited above the bandgap (**Supplementary Fig. 17c**), indicating that they arise from coupling between photoinduced charge-carriers and longitudinal optical (LO) phonon modes (*i.e.*, Fröhlich coupling). By fitting the decay in the photo-induced absorption (PIA) and taking a Fast Fourier Transform of the damped oscillations, we found three vibrational modes to appear (**Fig. 4b**, bottom panel). These vibrational modes match well with the Raman spectrum, measured from the same NC sample, consistent with these being LO phonon modes. By matching the energy of these vibrational modes with the calculated phonon dispersion curve for BiSBr (**Supplementary Note 8** and **Supplementary Fig. 20a**), we assigned their symmetries to $B_{2g}$, $B_{1g}$, and $A_g$ (**Fig. 4b** insets). We propose that the stronger oscillations for NCs compared to bulk thin films is attributed to their higher concentration of defect clusters that 1) provide more space for atomic displacements (Fig. 1f), enabling greater population of LO phonon modes, and 2) introduce extra vibrational phonon modes, which is consistent with our analysis of the phonon dispersion curves from first-principles calculations (**Supplementary Fig. 20b**). The resulting stronger electron-phonon coupling could then increase the likelihood of hot carriers entering into small polaron states.

Low-temperature PL measurements from 7 K to 200 K (**Fig. 4c**) further support the conclusion that NC samples have more pronounced formation of hot small polaron states. In bulk samples with low defect density, the above-gap PL is completely quenched at 7 K and gradually appears with increasing temperature, suggesting that this above-gap PL is phonon-assisted with a certain energy barrier. In contrast, in NCs with higher defect density, the above-gap PL persists even at 7 K.

To semi-quantitatively quantify the impact of small defect-bound hot polarons on the population of cold carriers, we performed pump-push-probe TAS measurements (**Fig. 4d-e**, **Supplementary Note 9**). A 1300 nm wavelength push beam was applied 10 ps after the pump to re-excite trapped and band-edge carriers. We monitored the $\Delta A$ at the band-edge (1.88–1.91 eV) and at sub-gap trap states (1.75–1.77 eV). In bulk thin films, the push beam induced a $\Delta A$ increase of 136.3% at the band-edge and 183.0% at mid-gap defects. In NCs, the corresponding increases were both lower, at 78.6% and 141.9%, respectively. The reduced increase in $\Delta A$ for the NC samples is consistent with a higher density of hot small polaronic states in the NCs that capture carriers before they cool to the band-edge or mid-gap states (Fig. 4f), such that there are fewer cold-carriers available for re-excitation with the push beam.

**Discussion**

Based on these results, we propose a configuration coordinate diagram for hot small polaron formation in BiSBr (**Fig. 5**). After photoexcitation into a hot state, charge-carriers can relax to the band-edge to form cold-carriers, enter into a small polaron state (state 2), or be captured by sub-gap traps (state 1). This diagram is supported by PL measurements showing emission at, above and below the band-edge (**Fig. 3a-b**), which can arise when photoexcited BiSBr couples to the ground state from states 1, 2 or from the cold excited state. This is also consistent with the global fitting to the TAS data (**Supplementary Fig. 18**), which requires a model with three relaxation states for hot carriers.

We propose that when there is a higher concentration of traps with transition levels in the conduction band, it is easier for hot carriers to enter into state 2. These hot carriers may first be captured by defect states at these above-gap transition levels, followed by local lattice relaxation to form small polarons. Alternatively, the formation of defect clusters could enhance the population of phonon modes, or introduce new phonon modes that result in stronger electron-phonon coupling, facilitating small polaron formation. These two processes are not mutually exclusive and may in fact be interrelated, with defect clustering both creating trapping centres and modifying the lattice dynamics that enable carrier localization. Consequently, NCs with more vacancy clusters readily small defect-bound hot polarons. This results in faster photoconductivity decay in OPTP, as well as stronger above-gap PL signals that persist even down to 7 K. These proposed mechanisms are consistent with the pump-push-probe TAS measurements, which indicate that there are fewer carriers available for re-excitation from the cold state or state 1 (**Fig. 4d-f**) because more of these carriers are held in state 2. Furthermore, photoluminescence excitation (PLE) measurements (**Supplementary Fig. 14**) suggests that the band-edge PL becomes stronger when the pump laser wavelength approaches the bandgap. This is consistent with there being fewer hot carriers localized in state 2 if we are exciting below the energy of state 2.

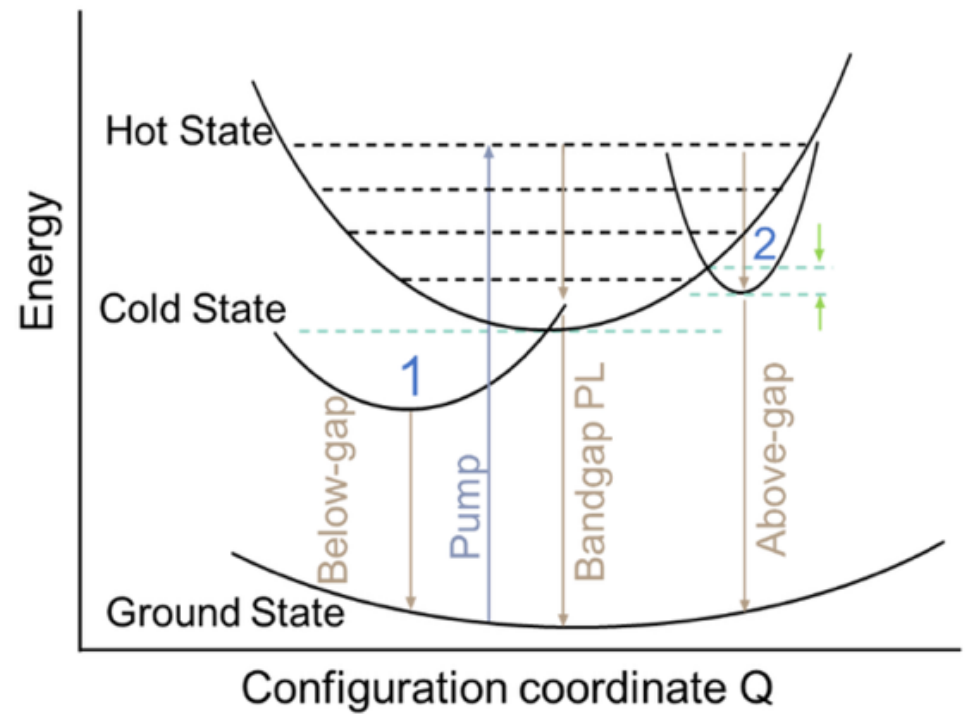

**Fig. 5 | Proposed configuration coordinate diagram of BiSBr.** In this proposed mechanism, photoexcitation causes the system to transition from the ground to hot state. The hot carriers can then relax to the band edge (causing the system to relax to the excited state), relax to a defect state (causing the system to enter into state 1), or enter into a hot small polaron state through coupling with vibrational states introduced through defects (causing the system to enter into state 2).

In summary, this work shows that carrier localisation in pnictogen-based PIMs is not determined solely by intrinsic lattice properties, but could also arise through defect-bound polarons. In the calculated defect formation energy diagram of BiSBr as a function of the Fermi level in **Fig. 2**, several defect transition levels appear above the bandgap. Having transition levels occurring outside the bandgap is commonly found, but what is unusual in the case of BiSBr is that these transition levels are not benign, as typically assumed, but play an active role in carrier localization of hot carriers. The coexistence of above-gap defect states and moderate polaron binding energies makes BiSBr an ideal system for disentangling the interplay between defects, carrier localisation, and hot-carrier dynamics. By combining first-principles calculations with advanced spectroscopic measurements, we show that BiSBr intrinsically supports band-like transport due to its high electronic dimensionality and low acoustic deformation potentials. In contrast, introducing point defects or defect clusters create local vibrational modes that couple strongly to excited carriers, leading to the formation of defect-bound hot polarons. These states intercept hot carriers during cooling, giving rise to extrinsic carrier localisation and a substantial reduction in the mobile carrier population.

By identifying defect-bound hot polarons as a key mechanism for carrier localisation, this work clarifies why experimental carrier mobilities in pnictogen-based materials often fall short of intrinsic predictions. The findings shift the view that self-trapping is an inherent feature of perovskite-inspired materials, and instead emphasise that defect engineering is not only critical for reducing non-radiative recombination, but also affects carrier localisation. This establishes a mechanistic basis for designing pnictogen-based absorbers with improved transport properties and provides guidance for developing next generation perovskite-inspired semiconductors for high performance optoelectronics.

**Acknowledgements**

X. G., J. Y. and R. L. Z. H. thank the UK Research and Innovation for funding through a Frontier Grant, awarded via the 2021 ERC Starting Grant scheme (no. EP/X029900/1). J. Y. and R. L. Z. H. also thank St. John's College, Oxford for support through the Welcome and Large grants. Y.-T. H. and R. L. Z. H. thank the Engineering and Physical Sciences Research Council (EPSRC) for funding (no. EP/V014498/2). C.L, E.S and C.C. acknowledge support from the Maria de Maeztu Units of Excellence Programme CEX2023-001300-M funded by MCIN/AEI (10.13039/501100011033). The authors would like to acknowledge the University of Warwick Research Technology Platform, Warwick Centre for Ultrafast Spectroscopy, for use of the optical pump terahertz probe spectrometer in the research described in this paper. Parts of this research were carried out at ELBE at the Helmholtz-Zentrum Dresden – Rossendorf e. V., a member of the Helmholtz Association. We would like to thank the facility staff for assistance. This work was partially supported by the Initiative and Networking Fund of the Helmholtz Association (FKZ VH-VI-442 Memriox), and the Helmholtz Energy Materials Characterization Platform (03ET7015). C. L. acknowledges support from the Spanish Ministry of Science, Innovation and Universities under an FPU grant. C.C. acknowledges support by MICIN/AEI/10.13039/501100011033 and ERDF/EU under the grants PID2023-146623NB-I00, PID2023-147469NB-C21 and CEX2023-001300-M, and by the Generalitat de Catalunya under the grants 2021SGR-00343, 2021SGR-01519 and 2021SGR-01411. Computational support was provided by the Red Española de Supercomputación under the grants FI-2024-3-0004, FI-2025-1-0007 and FI-2025-2-0006. B. X. thanks the Cambridge Trust, the China Scholarship Council and the Winton Programme for the Physics of Sustainability for funding. Z. Z. acknowledges supports from the National Natural Science Foundation of China (No. 12474031 and W2433157), Guangdong province (2025A1515010298), Guangzhou city (SL2023A04J00824, 2024A04J3731) and the GJYC program (2024D01J0078, 2024D03J0007). Y.B. acknowledges the Winton Programme for Physics of Sustainability for funding. R. L. Z. H. thanks the Science & Technology Facilities Council and Royal Academy of Engineering for support through the Senior Research Fellowships scheme (no. RCSRF2324-18-68). E.S. acknowledge the European Union's Horizon research and innovation programme under grant agreements number 866018 (SENSATE), the Spanish Ministry of Science and Innovation project number PID2023-148976OB-C41 (CURIO-CITY), and the ICREA Academia Program.

**Author contributions**

X. G., J. Y., and R. L. Z. H. conceived of the project. X. G. synthesized the bulk and NC BiSBr thin films with assistance from J. Y. X. G. conducted X-ray diffraction and Raman measurements. J. Y. performed pump-probe transient absorption measurements with support from B. X. J. Y. also conducted photoluminescence spectroscopy under different excitation conditions as well as temperature-dependent photoluminescence measurements with support from M. A. J. Y. performed time-revolved photoluminescence spectroscopy with assistance from Q. W. under the supervision of R.A.T. C.L.A. carried out DFT calculations under the supervision of E. S. and C. C., with support from S. R. K. and H. L. M. O .L. performed positron annihilation spectroscopy measurements with the support from A. W. and E. H., and M. B. carried out the ATSUP calculations. J. W. conducted pump-push-probe transient absorption measurements under the supervision of Z. Z. Y. H. performed optical pump terahertz probe measurements. L. V. T. and Y. B. carried out photothermal deflection spectroscopy measurements under the supervision of A. R. R. L. Z. H. supervised the overall work. All authors contributed to the writing and editing of the manuscript.

**Conflicts of Interest**

The authors declare no competing interests.

**Data Availability**

Raw experimental and computational data is available from the Oxford Research Archive [URL to be included before publication]